\documentclass[iop,apjl,numberedappendix,twocolappendix]{emulateapj}
\newcommand{\lea}{{\>\rlap{\raise2pt\hbox{$<$}}\lower3pt\hbox{$\sim$} \>}}
\newcommand{\gea}{{\>\rlap{\raise2pt\hbox{$>$}}\lower3pt\hbox{$\sim$} \>}}

\def\kms{km\,s$^{-1}$}
\def\rh{$r_{\rm h}$}

\slugcomment{Astrophysical Journal Letters, published 23 July 2015} 
\shorttitle{Dense stellar systems in SDSS}
\shortauthors{Sandoval et al.}

\begin{document}

\title{Hiding in plain sight: record-breaking compact stellar systems \\  in the Sloan Digital Sky Survey}

\author{Michael A. Sandoval\altaffilmark{1}, Richard P.\ Vo\altaffilmark{1,2},
              Aaron J.\ Romanowsky\altaffilmark{1,3},
Jay Strader\altaffilmark{4}, 
Jieun Choi\altaffilmark{5,6},  \\
Zachary G. Jennings\altaffilmark{6},  
Charlie Conroy\altaffilmark{5},  
Jean P.\ Brodie\altaffilmark{3},  
Caroline Foster\altaffilmark{7}, 
Alexa Villaume\altaffilmark{6}, \\
Mark A.\ Norris\altaffilmark{8}, Joachim Janz\altaffilmark{9},  
Duncan A.\ Forbes\altaffilmark{9}  
              }

\altaffiltext{1}{Department of Physics and Astronomy, San Jos{\'e} State University,  One Washington Square, San Jose, CA 95192, USA}
\altaffiltext{2}{Department of Physics and Astronomy, San Francisco State University, San Francisco, CA, 94131, USA} 
\altaffiltext{3}{University of California Observatories, 1156 High Street, Santa Cruz, CA 95064, USA}
\altaffiltext{4}{Department of Physics and Astronomy, Michigan State University, East Lansing, MI 48824, USA}
\altaffiltext{5}{Harvard--Smithsonian Center for Astrophysics, 60 Garden Street, Cambridge, MA 02138, USA}
\altaffiltext{6}{Department of Astronomy and Astrophysics, University of California, Santa Cruz, CA 95064, USA}
\altaffiltext{7}{Australian Astronomical Observatory, P.O. Box 915, North Ryde, NSW 1670, Australia}
\altaffiltext{8}{Max Planck Institut f\"ur Astronomie, K\"onigstuhl 17, D-69117, Heidelberg, Germany}
\altaffiltext{9}{Centre for Astrophysics \& Supercomputing, Swinburne University, Hawthorn, VIC 3122, Australia}

\begin{abstract}
Motivated by the recent, serendipitous discovery of the densest known galaxy, M60-UCD1, we present
two initial findings from a follow-up search, using the Sloan Digital Sky Survey, Subaru/Suprime-Cam
and {\it Hubble Space Telescope} imaging, and SOuthern Astrophysical Research (SOAR)/Goodman spectroscopy.
The first object discovered, M59-UCD3, has a similar size to M60-UCD1 (half-light radius of $r_{\rm h}$\,$\sim$\,20\,pc) but
is 40\% more luminous ($M_V \sim -14.6$), making it the new densest-known galaxy. 
The second, M85-HCC1, has a size like a typical globular cluster (GC; $r_{\rm h}$\,$\sim$\,1.8\,pc) but is much more luminous
($M_V \sim -12.5$). 
This hypercompact cluster is by far the densest confirmed free-floating stellar system, and is equivalent to the densest known nuclear star clusters.
From spectroscopy, we find that both objects are 
relatively young ($\sim$\,9\,Gyr and $\sim$\,3\,Gyr, respectively), with metal-abundances that resemble those of galaxy centers.
Their host galaxies show clear signs of large-scale disturbances, and we conclude that these dense objects are the
remnant nuclei of recently accreted galaxies.
M59-UCD3 is an ideal target for follow-up with high-resolution imaging and spectroscopy to 
search for an overweight central supermassive black hole as was discovered in M60-UCD1.
These findings also emphasize the potential value of
ultra-compact dwarfs and massive GCs as tracers of the assembly histories of galaxies.
\end{abstract}


\section{Introduction}

The classic distinction between galaxies and star clusters was riven by the discovery of stellar systems with
intermediate sizes and luminosities: the ultracompact dwarfs (UCDs; \citealt{Hilker99,Drinkwater00}). 
The nature and origins of these novel objects have been debated ever since, with potentially important implications
for how star clusters and galaxies form and evolve---tracing
novel modes of star formation, cluster merging, and/or episodes of satellite galaxy accretion
(e.g., \citealt{Fellhauer02,Pfeffer14}).

The UCDs were previously overlooked, not because they were extremely rare, nor 
especially difficult to observe, but because they did not fit 
in with preconceptions about known object types.
They were, therefore, filtered out during the focused search process (cf.\ similar oversights discussed in \citealt{Simons99,Drew13}). 
To those using ground-based imaging to study extragalactic globular clusters (GCs), the UCDs 
were deemed too bright, and assumed to be foreground stars.  To those using the {\it Hubble Space Telescope} ({\it HST}), 
whose fine spatial resolution is well suited for appreciating the extended nature of the UCDs,
they appeared too diffuse, and were seen as background galaxies.

Despite this lesson in selection bias, years of research on UCDs ensued without questioning whether or not the
parameter space of their properties had been adequately mapped out.
The impact of this shortcoming was exemplified by the emphasis on an apparent size--luminosity relation for UCDs
(e.g., \citealt{Kissler06,Murray09,Gieles10}),  
which was later argued to be merely a consequence of observational limitations at low surface-brightnesses,
with the population of large UCDs actually extending to much lower luminosities \citep{Brodie11}.

It was also assumed that UCDs were restricted to high-density environments, as they
were first identified around the central galaxies in the Fornax and Virgo clusters, and indeed had earlier been predicted
to form in this context \citep{Bassino94}.
However, UCDs were subsequently found around ordinary field galaxies, implying that their formation does
not require such particular circumstances (\citealt{Hau09,Norris11}; \citealt{Norris14}, hereafter N+14).

As a recent step toward  a broader understanding of UCDs and other compact stellar systems,
\citet{Strader12} analyzed a mosaic of {\it HST}/Advanced Camera for Surveys (ACS) images of the Virgo giant elliptical galaxy M60 (NGC~4649),
and scrutinized {\it all} the detected objects to consider whether they might be associated with M60 rather than
being discardable as foreground or background contaminants.
Consequently, in this single galaxy, spectroscopic follow-up revealed two new varieties of UCDs:
very low-luminosity, diffuse objects that bridged the gap between UCDs and extended star clusters \citep{Forbes13},
and the densest galaxy yet discovered, M60-UCD1 \citep[hereafter S+13]{Strader13}.

M60-UCD1 has a luminosity of $M_V = -14.2$, a stellar mass of $M_\star \simeq 2\times10^8 M_\odot$,
a half-light radius of $r_{\rm h} \simeq 25$\,pc, and a velocity dispersion of $\sigma \simeq 70$\,\kms---properties
that are intermediate between the classical UCDs and the compact elliptical galaxies (cEs). 
This area of parameter space is being studied by the Archive of Intermediate Mass Stellar Systems
(AIMSS) Project, which uses archival {\it HST} images in the first systematic survey for UCDs across the sky (N+14).
AIMSS has so far yielded the independent discovery of M60-UCD1 along with somewhat larger objects that
now fully bridge the gap between UCDs and cEs and establish a firm link between these seemingly disparate families of stellar systems.

The UCD--cE transition objects have taken on a new dimension of importance with the discovery of a supermassive
black hole (SMBH) in M60-UCD1 that accounts for a remarkable 15\% of its host galaxy mass \citep{Seth14}.
These ``overweight'' SMBHs may be a pervasive phenemonon among UCDs, based on their elevated mass-to-light ratios
\citep{Mieske13,Forbes14,Janz15}.
It is therefore timely to search for new objects in this class, some of which have
been hiding in plain sight for decades, as they are bright enough to be visible on 
Digitized Sky Survey images and are included as ``stars'' in the USNO catalogs.
For example, M60-UCD1 has an apparent magnitude of $r = 16.7$, is visible in early photographs
\citep{Hubble22},
and could easily have been discovered through serendipitous spectroscopy,
had it been a target in the Sloan Digital Sky Survey (SDSS),\footnote{\tt http://www.sdss.org} as was its fainter cousin M59cO  \citep{Chilingarian08}.

One limitation of AIMSS is its confinement to the narrow  {\it HST} camera footprints, covering out to
$\sim$\,2$^{\prime}$ in galactocentric radius ($\sim$~10~kpc in Virgo).
The natural venue for continuing the search to larger radii and across a wide swath of the sky is the SDSS,
which is the basis for our new UCD data-mining program, with initial results reported here.
These include the identification of remarkable, compact stellar systems around two Virgo galaxies:
M59 (NGC~4621) and M85 (NGC~4382), at distances of $14.9\pm0.5$ and $17.9\pm0.5$\,Mpc, respectively \citep{Blakeslee09}.
The discoveries of these objects are presented in Section~\ref{sec:search},
the implications discussed in Section~\ref{sec:context},
and a summary provided in Section~\ref{sec:end}.

\section{The Search}\label{sec:search}

We have used fairly simple methods to carry out a very rapid search for novel objects in SDSS, as we describe below,
with a photometric approach in Section~\ref{sec:phot} and 
a spectroscopic approach in Section~\ref{sec:spec}.

\subsection{Photometric Selection:  M59-UCD3}\label{sec:phot}

We began by using SDSS to search for objects similar to M60-UCD1 and M59cO around giant early-type galaxies (ETGs) in Virgo.
The defining characteristics were apparent magnitude ($V \sim 17$), color ($g-i \sim 1.2$), and size.
Although such objects are not resolved in SDSS imaging,
we noticed that they had ``GALAXY'' classifications, and  Petrosian radii of 2$^{\prime\prime}$--3$^{\prime\prime}$
(compared to $\sim$1.5$^{\prime\prime}$ for point sources), so their extended envelopes are detectable in 
ground-based imaging.

\begin{figure}
\centering{
\includegraphics[width=9.2cm]{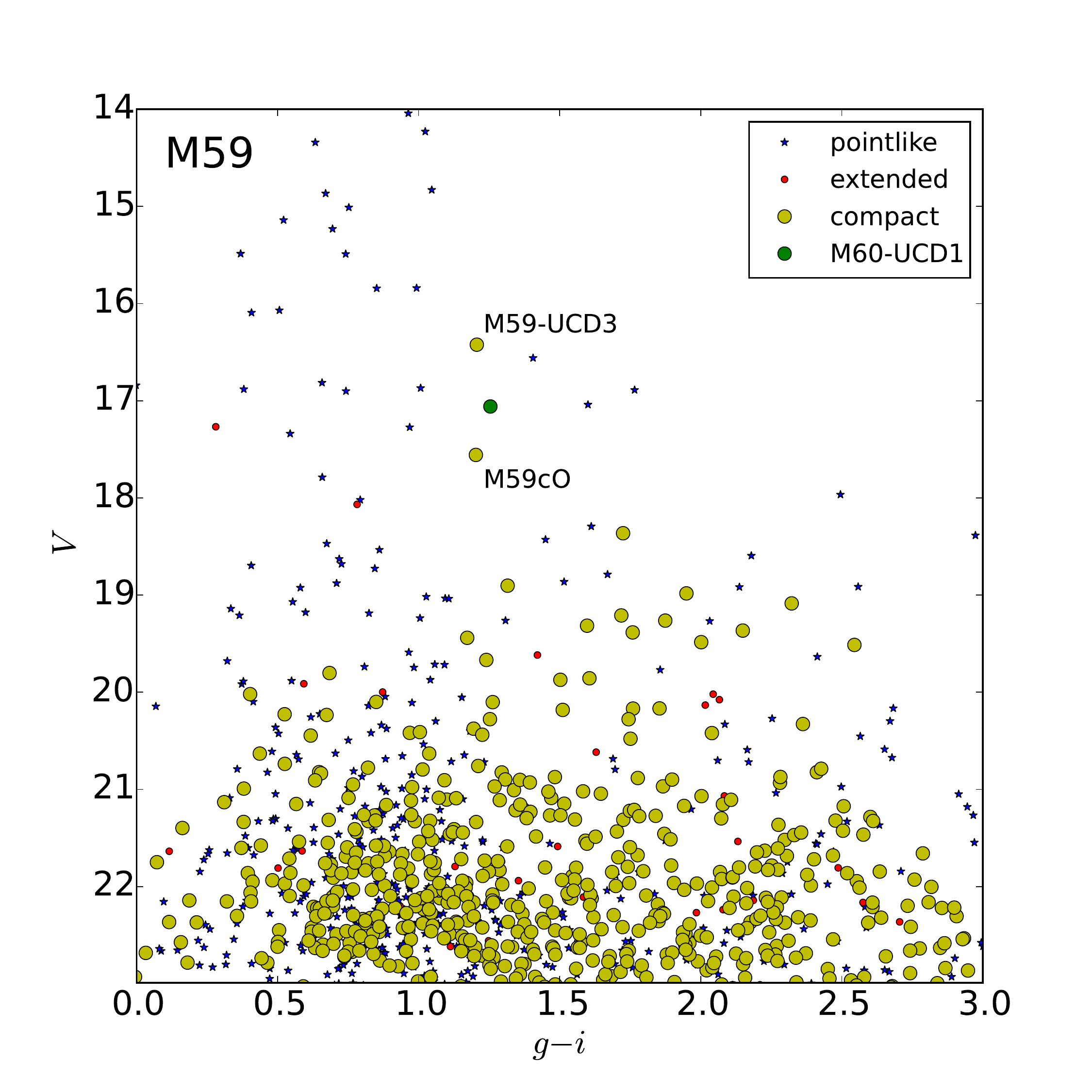} 
}
\caption{Color--magnitude diagram of sources around M59, based on SDSS-DR7 photometry \citep{Abazajian09}, 
out to a galactocentric radius of 7.1$^{\prime}$ (10 galaxy effective radii).
The $V$-band magnitudes are calculated using $g-r$ color conversions from 
SDSS.\footnote{\tt http://www.sdss.org/dr12/algorithms/sdssUBVRITransform/}
Symbols denote object classifications in the legend, where pointlike objects are ``stars,'' while extended and compact objects are
``galaxies'' with Petrosian radii greater than and less than 4$^{\prime\prime}$, respectively.
M60-UCD1 is also marked.  Two comparable objects appear at $V \sim 17$: 
M59cO, and a new object, M59-UCD3.
}\label{fig:cmd}
\end{figure}

\begin{figure*}
\centering{
\includegraphics[width=16.5cm]{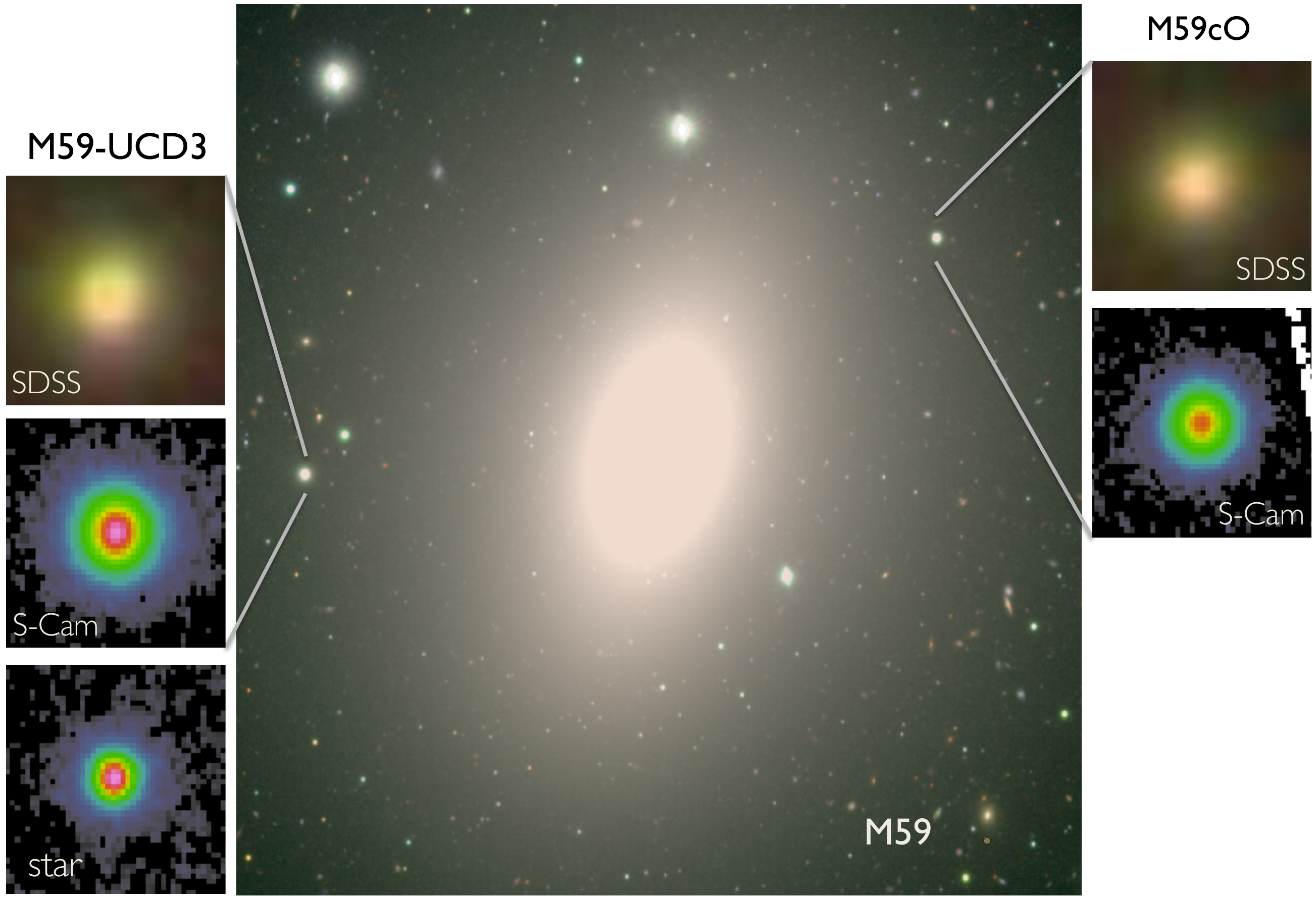} 
\vskip 0.2cm
\includegraphics[width=16.3cm]{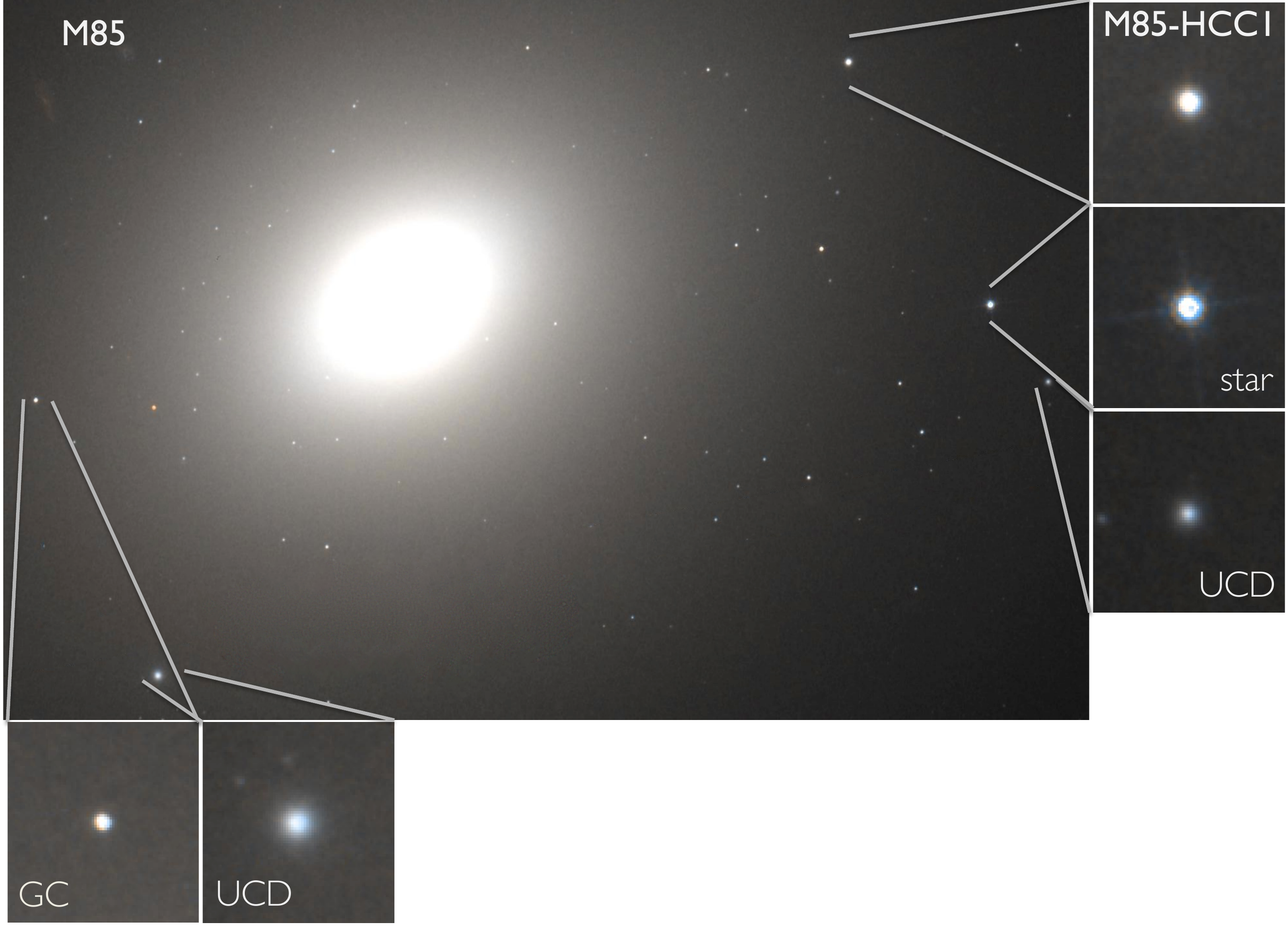}
}
\caption{{\it Top:} M59 and its accompanying UCDs.
The central panel is a $\sim 5^{\prime}$ (22\,kpc) square Suprime-Cam color image. 
At top on left and right are 8$^{\prime\prime}$ (600\,pc) SDSS zoom-ins of M59-UCD3 and M59cO, respectively.
Underneath are $i$-band thumbnails from Subaru/Suprime-Cam, using false color for contrast.
At bottom-left is a comparison star, using the same scalings.
M59-UCD3 is visibly more extended than the star.
{\it Bottom:} {\it HST}/ACS image of M85 and surrounding objects, as marked,
covering $\sim1.4^\prime \times 0.9^\prime$ ($7.3\times4.7$ kpc).
}\label{fig:images}
\end{figure*}

We started by calibrating our detection techniques on M60 and M59, given their known UCD-like objects.
Remarkably, the SDSS color--magnitude diagram of sources around M59 immediately showed 
M59cO plus another, similiar object (Figure~\ref{fig:cmd}). 
This second one, M59-UCD3, appears relatively close to M59
(2.2$^{\prime}$ or 9.3\,kpc projected distance; Figure~\ref{fig:images}), but was 10$^{\prime\prime}$ beyond the nearest
{\it HST} footprint.
Its photometric properties, from SDSS-DR10 \citep{Ahn14}, are listed in Table~\ref{tab:info}.
Assuming the M59 distance (confirmed below), the absolute magnitude is $M_V = -14.6$, i.e., $\sim$40\% more luminous than M60-UCD1. 

\begin{table}
\center{
\caption{Characteristics of compact stellar systems}\label{tab:info}
\begin{tabular}{ccc}
Parameter & M59-UCD3 & M85-HCC1 \\
\hline
R.A. [J2000] & 12:42:11.05 & 12:25:22.84 \\
Decl. [J2000] & $+$11:38:41.3  & $+$18:10:53.7  \\
$r_{\rm h}$ [pc] & $20\pm4$ & $1.85\pm0.9$ \\ 
$g$ & $16.81 \pm 0.05$ & $19.16 \pm 0.03$ \\ 
$r$ & $16.00 \pm 0.05$  & $18.55 \pm 0.05$ \\ 
$i$  & $15.61 \pm 0.05$  & $18.24 \pm 0.06$\\  
$V$ & $16.34 \pm 0.05$ & $18.80 \pm 0.03$ \\ 
$M_{V,0}$ & $-14.60 \pm 0.09$ & $-12.55 \pm 0.07$ \\ 
$v$ [km\,s$^{-1}$]  & $373 \pm 18$ & $658\pm4$ \\
$M_\star$ [$M_\odot$] & $(1.8\pm0.3)\times10^8$  & $(1.2\pm0.1)\times10^7$ \\ 
$\Sigma_{\rm h} \, [M_\odot$\,pc$^{-2}]$ & $7.1\times10^4$ & $5.8\times10^5$ \\ 
{} [Fe/H] [dex]  & $-0.01 \pm 0.04$ &$-0.06\pm0.07$ \\ 
{} [Mg/Fe] [dex] & $+0.13\pm0.07$ & $+0.05\pm0.13$ \\ 
{} mean age [Gyr] & $8.6\pm2.2$ & $3.0\pm0.4$ \\ 
\hline
\end{tabular}
}
\end{table}

M59 was imaged as a bonus galaxy in the Sages Legacy Unifying Globulars
and GalaxieS Survey (SLUGGS; \citealt{Brodie14}).\footnote{\tt http://sluggs.ucolick.org}
The SLUGGS ground-based imaging from Subaru/Suprime-Cam included both long and very short (10\,s) exposures,
to prevent saturation of bright GCs/UCDs.
We used $(g, r, i)$ filters,
with 0.6$^{\prime\prime}$--0.9$^{\prime\prime}$ seeing.

We reduced these images using a modified version of the SDFRED2 pipeline \citep{Ouchi04},
and found that 
M59-UCD3 is clearly extended when compared to a nearby bright star (Figure~\ref{fig:images}).
To measure the size, we use {\tt ishape} \citep{Larsen99}, which fits
 the two-dimensional image of the object with a model surface-brightness profile convolved 
with an empirical point-spread function (PSF).

Using the nearby star for PSF reference, we find a good fit with a King model, and a size of
$r_{\rm h}=0.28^{\prime\prime}$ or 20\,pc (with uncertainties dominated by systematics).
For a reliability check, we also analyze M59cO and M60-UCD1 from the same series of Suprime-Cam images,
and find sizes  $\sim 20\%$ lower than the {\it HST}-based measurements 
(\citealt{Chilingarian08}; S+13; N+14).
More robust size and profile information for M59-UCD3 will require {\it HST} follow-up,
which could reveal a two-component profile as in M59cO and M60-UCD1.

We obtained a spectrum of M59-UCD3 with
the Goodman High-Throughput Spectrograph \citep{Clemens04} on the 4.1\,m 
SOuthern Astrophysical Research (SOAR) telescope, on 2014 February 9. We
used a single 600~s exposure with a 400~l~mm$^{-1}$ grating and a 1.03$^{\prime\prime}$ slit, 
giving a wavelength coverage of $\sim$\,3100--7000\,\AA\ and a resolution of 5.7\,\AA. 
The spectrum was optimally extracted and wavelength calibrated in the standard manner using {\tt IRAF},
with skylines used to check for flexure,
and has a signal-to-noise ratio of S/N\,$\sim$\,55\,\AA$^{-1}$ 
(see top panel of Figure~\ref{fig:spec}).
The heliocentric recession-velocity of $(373\pm18)$\,\kms\
was derived through cross-correlation with an early K-giant spectrum  taken with the same setup.
This velocity differs by $\sim -90$\,\kms\ from the center of M59, further
indicating an association between this large galaxy and M59-UCD3.\footnote{During the late stages of paper preparation,
we learned that another group had independently identified M59-UCD3 as a compact galaxy candidate, 
using SDSS imaging,
and reported it at a conference \citep{Marzke06}.}

\subsection{Spectroscopic Selection: M85-HCC1}\label{sec:spec}

After identifying M59-UCD3, we continued searching for UCDs around other nearby, luminous ETGs.
In addition to SDSS imaging, we now incorporated information from
the SDSS spectroscopic database.
While far less complete than the photometry, the spectra, where available,
allow immediate categorization of many objects without the time-consuming
need for follow-up spectroscopy.
Such a ``Virtual Observatory'' approach was used to look for cEs
(e.g., \citealt{Chilingarian09,Huxor11}),
but to our knowledge there has been no systematic search for UCDs
via SDSS spectroscopy (the discovery of M59cO was serendipitous; \citealt{Chilingarian08}).

\begin{figure}
\centering{
\includegraphics[width=\columnwidth]{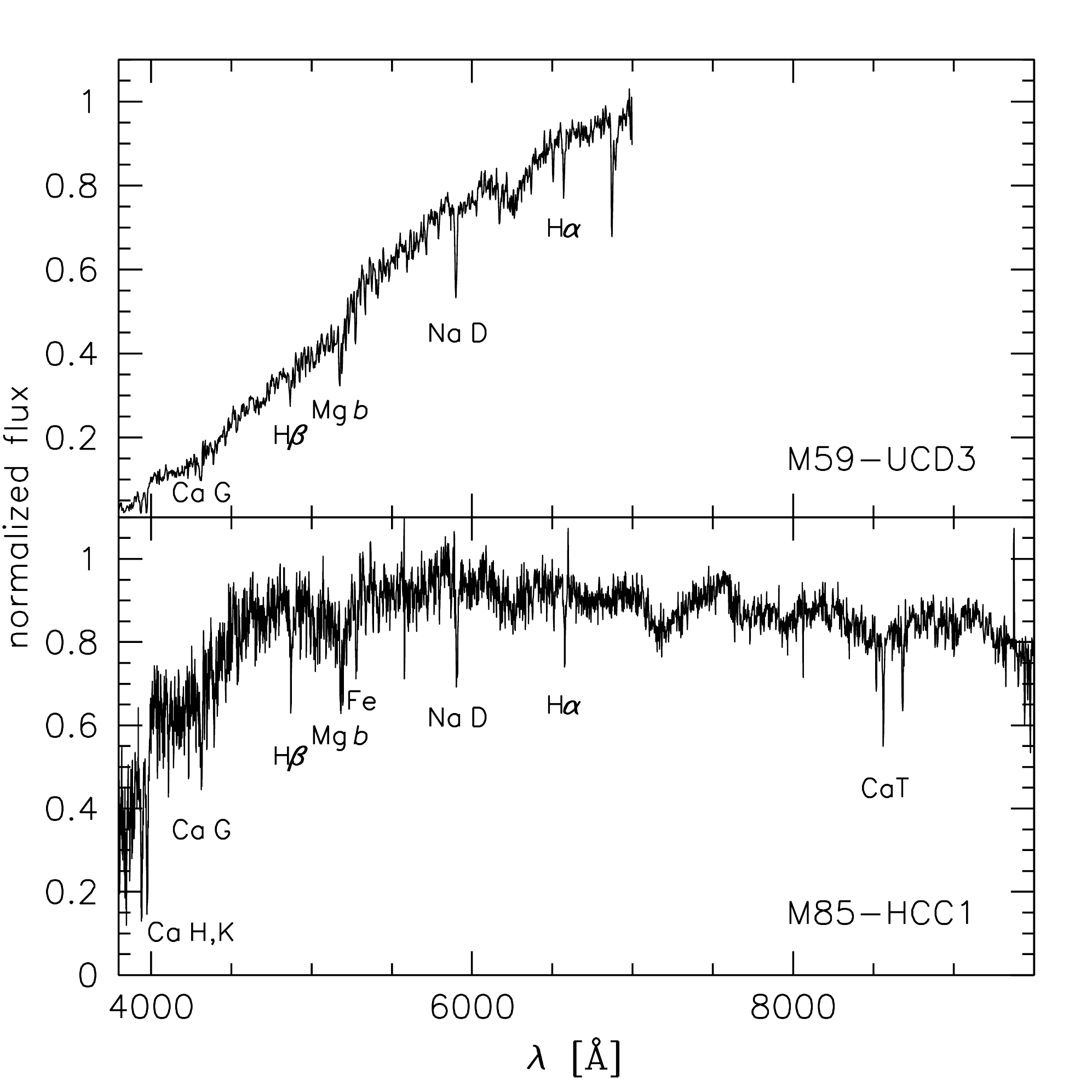}
}
\caption{Spectra of M59-UCD3 ({\it top}) and M85-HCC1 ({\it bottom}), from
SOAR/Goodman and SDSS-III/BOSS, respectively.
Key absorption lines are indicated.
}\label{fig:spec}
\end{figure}

Our spectroscopic search made use of SDSS-DR10, with its
expansion of the original SDSS spectral database as part of
 the SDSS-III surveys\footnote{\tt http://www.sdss3.org}
\citep{Dawson12}.
The first galaxy we checked was M85, a 
Virgo cluster lenticular with disturbed isophotes (we selected it because it was the most luminous
candidate not among the well-studied Virgo ellipticals such as M87 and M49).
We noticed an intriguing object adjacent to this galaxy (at 0.6$^{\prime}$ or 3.3\,kpc), 
which was classified as a ``STAR'' photometrically
(with a Petrosian radius of 1.4$^{\prime\prime}$), but with a spectrum classified as a ``GALAXY'', and a
recession velocity of $658\pm4$\,km\,s$^{-1}$---too high for a normal Galactic star, and 
similar to the velocity of M85 ($\simeq$\,730\,km\,s$^{-1}$).

This object was also included in an {\it HST}/ACS image (Figure~\ref{fig:images}).
Here it appeared very bright and compact but with much less of a diffraction-spike than obvious stars of similar brightness.
Our initial {\tt ishape} analysis, using a few PSF-reference stars, yielded a barely resolved
size of $\sim$\,1--2\,pc, depending on the detailed fit-parameters.

A literature search revealed that the object was previously reported as a photometric GC-candidate 
with $r_{\rm h}$ of 1.8--1.9\,pc, after detailed mapping of PSF variations
(\citealt{Jordan09,Chies11}).\footnote{We find that standard King models are not particularly good
fits to the existing imaging, and estimate an uncertainty of
$\sim$50\% in $r_{\rm h}$, which should be revisited with future data.}
This was the brightest GC candidate cataloged by the ACS Virgo Cluster Survey, just barely
making it past their $z\geq18.0$ selection limit \citep{Jordan04}. 

As discussed in the next section, the properties of this object are unusual enough to motivate
a new term, the ``hypercompact cluster," and we have named it M85-HCC1.  Its spectrum is shown in Figure~\ref{fig:spec} (bottom panel).

\section{The new objects in context}\label{sec:context}

After confirming the new objects, and measuring some basic properties, 
here we present additional analyses, and discuss them in the context of other stellar systems.

\begin{figure*}
\centering{
\includegraphics[width=8.9cm]{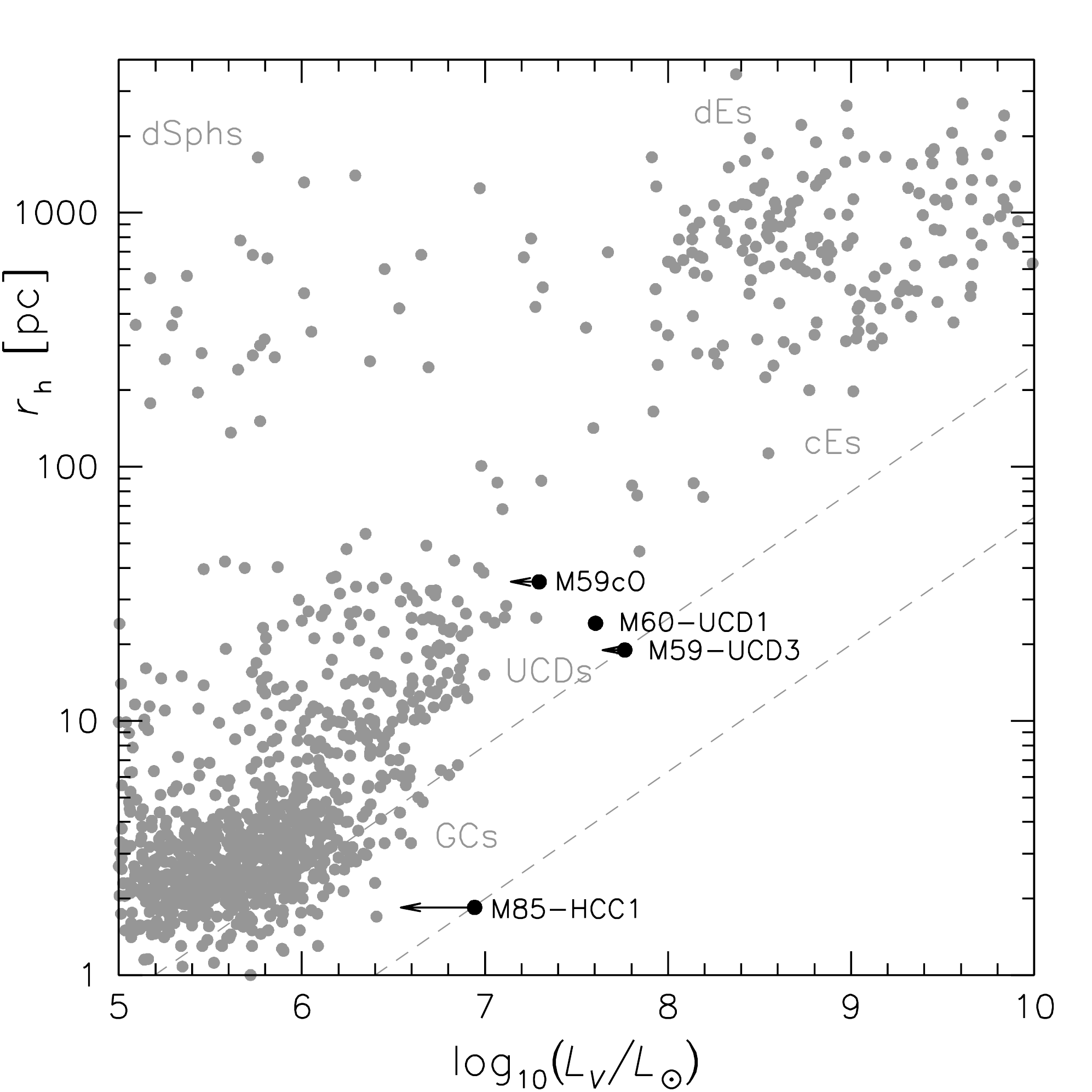} 
\includegraphics[width=8.9cm]{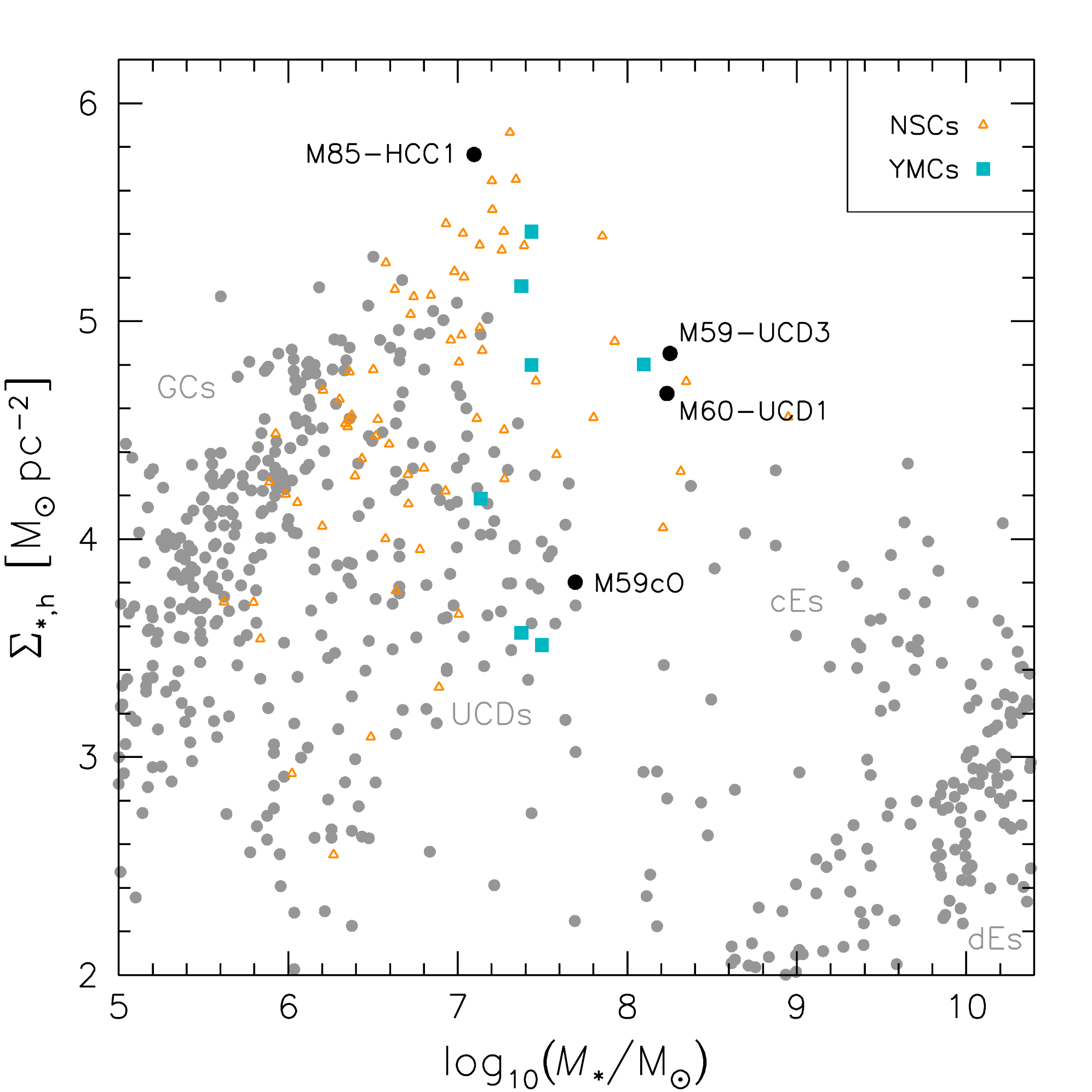} 
\includegraphics[width=8.95cm]{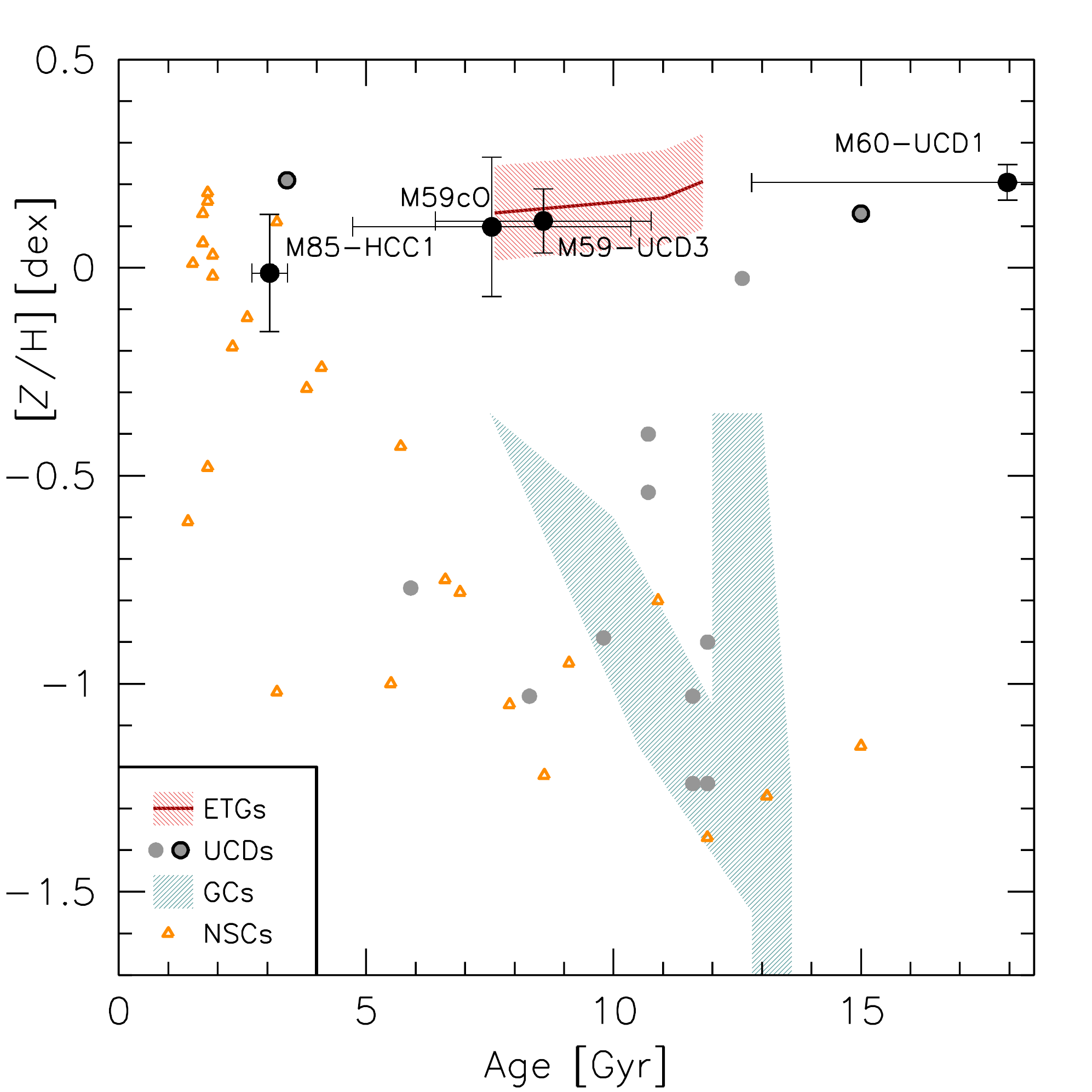}
\includegraphics[width=8.95cm]{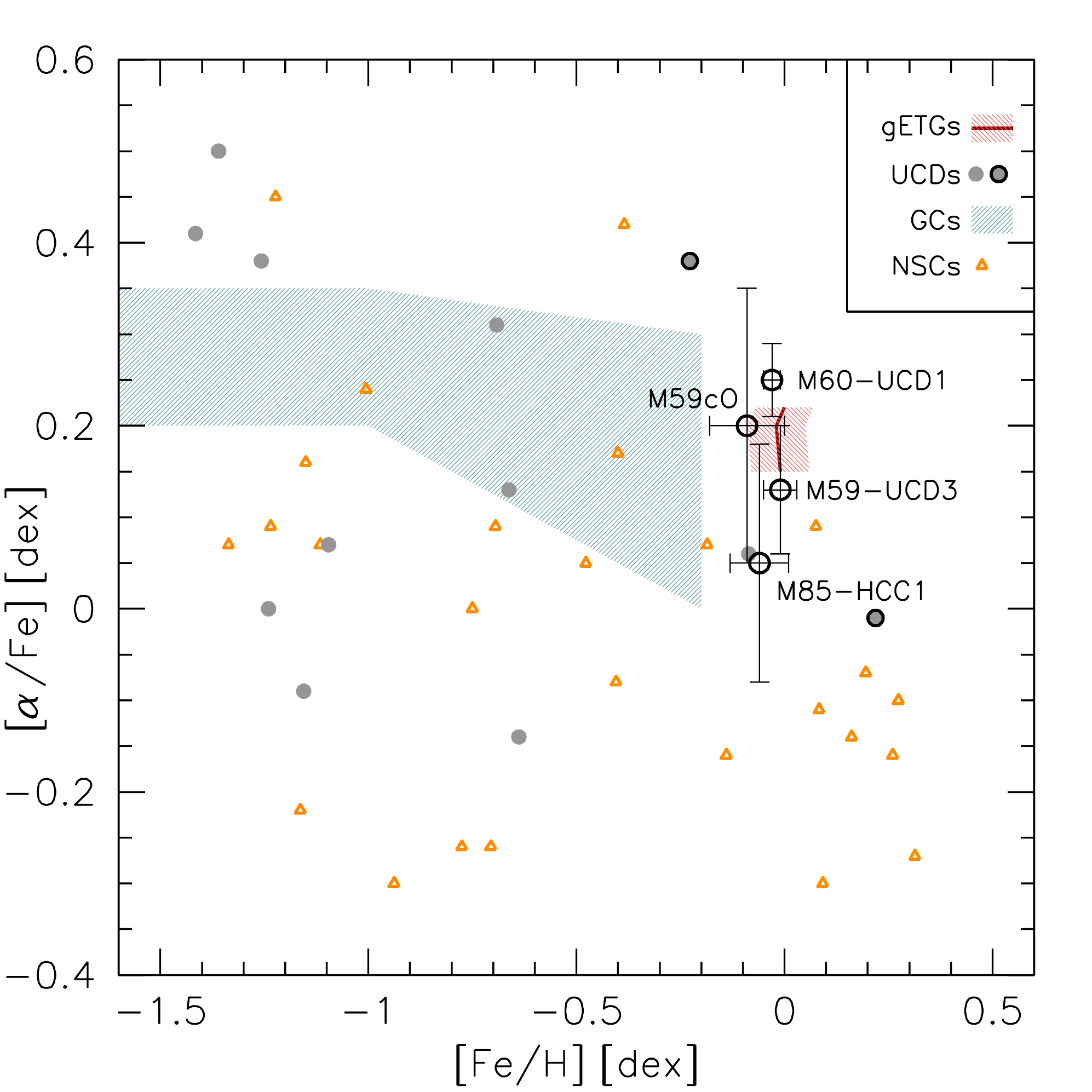}
}
\caption{Properties of newly discovered objects in relation to other stellar systems.
{\it Upper-left:} size--luminosity diagram of distance-confirmed stellar systems in the nearby universe.
Gray points are from a literature compilation
(originally from \citealt{Brodie11}, with updates maintained online at {\tt http://sages.ucolick.org/spectral\_database.html}),
and black points mark the four dense stellar systems discussed in this paper,
with arrows showing predictions for fading to 13-Gyr ages. 
The dashed lines show constant surface-brightness values of $2.5\times10^4 \, L_{V,\odot}$\,pc$^{-2}$ (top) and 
$4\times10^5 \, L_{V,\odot}$\,pc$^{-2}$ (bottom).
{\it Upper-right:} stellar average-density vs.\ mass (data from N+14).
{\it Lower panels:} stellar populations trends for the four dense stellar systems,
compared to other objects for context (see legends for symbols):
the centers of large ETGs, UCDs (black borders for $M_\star > 3\times10^7 M_\odot$), Milky Way GCs, and NSCs
(see \citealt{Pritzl05,Brodie11,Dotter11,Conroy14}; within the uncertainties, the oldest objects are consistent with 13~Gyr ages).
An origin is implied for the ``densest objects'' as the centers of galaxies.
}\label{fig:uber}
\end{figure*}

First, we place the objects in a diagram of size and luminosity for
nearby, spheroidal stellar systems, along with reference
lines of constant surface brightness  (Figure~\ref{fig:uber}, upper left).
For M59-UCD3, the surface brightness (averaged within \rh)
is $\sim 2.7\times10^4 \, L_{r,\odot}$\,pc$^{-2}$. 
This is twice as high as the previous galactic record-holder, M60-UCD1,
making M59-UCD3 a prime target for SMBH follow-up.
M59-UCD3 is a younger system (see below), reducing its stellar mass-to-light ratio, with
a stellar surface mass density of $\sim 7\times10^4 \, M_\odot$\,pc$^{-2}$
that is $\sim$50\% higher than that of M60-UCD1.
For volume densities, we use scalings from \citet{Wolf10},
and estimate $\sim 400 \, L_{r,\odot}$\,pc$^{-3}$ in luminosity,
and $\sim 1100 \, M_\odot$\,pc$^{-3}$ in mass.
Scaling to the few-thousand visible stars of the Solar neighborhood \citep{ESA97,Binney08}, an observer in the core of
M59-UCD3 would see around a million stars in the sky. 

Turning to M85-HCC1, 
its compactness combined with its luminosity places it nearly in a class of its own among confirmed, free-floating
old stellar systems, with a luminosity like the brightest classical UCDs but with a size like the smallest GCs 
(Figure~\ref{fig:uber}).
Although fading with age would place it close to the previously densest-known star cluster, NGC~4494-UCD1 \citep{Foster11},
that object also shows signs of fading youth \citep{Usher15}.
The mean surface brightness and mass density of M85-HCC1 
are $\sim 4\times10^5 L_{r,\odot}$\,pc$^{-2}$ and $\sim 6\times10^5 M_\odot$\,pc$^{-2}$,
which is $\sim10\times$ denser than M59-UCD3.  The volume densities are $\sim100\times$ higher:
$\sim 7\times10^4 L_{r,\odot}$\,pc$^{-3}$ and $\sim 10^5 M_\odot$\,pc$^{-3}$.

For additional context, we compare to  two classes of object with possible evolutionary
connections to UCDs and GCs: young massive clusters (YMCs,
generally formed during galactic starbursts),
and galactic nuclei---also called nuclear star clusters (NSCs).
Stellar masses should be compared, rather than luminosities, since
young objects will naturally fade with time (a key issue for the YMCs and NSCs, not so much for the GCs/UCDs).

For our new objects' stellar masses, we use spectral model fitting (below).
For YMCs and NSCs, N+14 have provided a compilation.
These objects have a wide range of densities, making them plausible as progenitors
of both GCs and UCDs. However, the maximum reported YMC density is $2.5\times10^5 M_\odot$\,pc$^{-2}$---lower
than M85-HCC1 by half.  NSCs, on the other hand, reach higher densities, with the densest one
(in the low-luminosity ETG NGC~4476)  even surpassing M85-HCC1.
We thus conclude that M85-HCC1 is most likely  a tidally stripped galaxy-center, and
also argue the same for M59-UCD3 based on its properties that are closely analogous to M60-UCD1.

More evidence for the galactic-center origins of both objects would come if overweight SMBHs were detected, as in M60-UCD1.
We note that the original motivation for SDSS-III spectroscopy of M85-HCC1 was its X-ray source, 
which has fairly soft emission and a luminosity of $L_{\rm X} = 2.6\times10^{38}$\,erg\,s$^{-1}$ \citep{Sivakoff03}.
However, this emission might simply reflect a low-mass X-ray binary, as
the estimated stellar encounter rate is extremely high.

We turn next to clues from stellar populations, using stellar population synthesis with full spectral fitting and
variable abundance ratios, as in \citet{Conroy14} and \citet{Choi14}.
We fit over the wavelength range 4050--6550\,\AA, using a high-order polynomial to normalize the continuum,
and a Markov Chain Monte Carlo algorithm to explore parameter space and return uncertainties.

The resulting ages, metallicities, and alpha-element enhancements (traced by [Mg/Fe]) are reported in Table~\ref{tab:info}.
Both objects have near-solar metallicities, and
probably mildly elevated [Mg/Fe], implying fairly rapid timescales of star formation.
Their inferred ages are relatively young: $\sim$6--11\,Gyr for M59-UCD3, and $\sim 3$\,Gyr for M85-HCC1.
For comparison, we have re-modeled the available spectra from M60-UCD1 (MMT/Hectospec) and M59cO (SDSS),
and placed them in plots of metallicity versus age and [Mg/Fe] (Figure~\ref{fig:uber}, lower-panels).
M59-UCD3 shows similar properties to M59cO, suggesting they may be coeval
(perhaps arising from the same or associated merger event).

For additional context, we also plot other objects from the literature.
M59cO, M59-UCD3, and M60-UCD1 follow a high-metallicity trend seen in other massive UCDs
(as discussed in \citealt{Brodie11}),  and
in the centers of large ETGs---in contrast to dwarf-galaxy nuclei and to GCs.
M85-HCC1 is an ambiguous case, and might have originated in either a dwarf or giant galaxy.
A more detailed discussion of individual elements is postponed, while here we only note an intriguing finding that nitrogen and sodium
are elevated in M59cO and M60-UCD1 (see also S+13), but not in M59-UCD3 and M85-HCC1.

We also consider the current environments of these four unusual objects.
M60 shows indications of substructure in its GC system \citep{DAbrusco15}, although
M60-UCD1 may have formed in a different, earlier event.
The host galaxy M85 is well known to show large-scale disturbances in its stellar light, signifying a merger  within the
past few Gyr, which is also reflected in a $\sim$\,1.5--2.0\,Gyr age for both its central stars and the bulk of its central GCs
\citep{Trancho14}.  It is furthermore the ``poster-child'' for red, diffuse star clusters, with $\sim$\,160 of these curious objects within
the central $\sim$\,10\,kpc \citep{Peng06}. These are generally thought to be residues either of gas-rich mergers or of faded open clusters from 
spiral disks.

M59 is a less apparent merger case, 
 as may be expected given the older ages of M59-UCD3 and M59cO.
However, it hosts a dramatic, coherent central GC substructure \citep{DAbrusco15}.
We speculate based on all the available information that M59 and M85 experienced fairly recent mergers with
intermediate-mass early-type and late-type galaxies, respectively.
\\

\section{Conclusions}\label{sec:end}

We have presented the discoveries of two unique compact stellar systems, M59-UCD3 and M85-HCC1,
based initially on publicly available photometric and spectroscopic data from SDSS/SDSS-III and {\it HST}.
Further confirmation and characterization have come from Subaru imaging and SOAR spectroscopy.

These two objects were found to be the densest galaxy and densest free-floating star system, respectively---raising questions about
how many more such objects exist, and how well the boundaries for stellar systems in size--luminosity parameter space are known.
The objects may also provide clues to the physics underlying the observed maximum central surface density for stellar
systems of $\sim10^5 M_\odot$\,pc$^{-2}$ \citep{Hopkins10}.

Through examination of density, age, metallicity, alpha-elements, and X-ray emission, along with evidence for recent mergers in the
host galaxies, we conclude that the two new objects, plus two from the literature, are most likely the stripped centers of galaxies.
There is strong potential for confirming this scenario in M59-UCD3 by adaptive-optics search for an overmassive SMBH.
We also emphasize that the post-merger indestructibility
of such dense objects may in general allow them to be used as tags for the assembly histories of their host galaxies.

\acknowledgements

We thank the SJSU ASTR-117B class for their endurance through UCD fever, Katy Murphy for helpful discussions,
Cindy Tsui for artistic assistance, and the referee for a quick and helpful report.
We are immensely grateful for the efforts of SDSS, SDSS-III, and the Hubble Legacy Archive
in making their data easily accessible to anyone in the world with an internet connection.
This work was supported by National Science Foundation (NSF) grant AST-1109878,
by SJSU Undergraduate Research Grants,
and by the ARC (DP130100388).
It used data from the Subaru telescope
(National Astronomical Observatory of Japan) and  
the Southern Astrophysical Research (SOAR) telescope (joint project of Brasil, NOAO, UNC, MSU).
SDSS/SDSS-III was funded by the Alfred P.\ Sloan Foundation, the
Participating Institutions, the NSF,  the U.S.\ Department of 
Energy, the National Aeronautics and Space Administration, the Japanese Monbukagakusho, the Max Planck Society,
and the Higher Education Funding Council for England.

\end{document}